# ESTIMATION OF MEDIUM ACCESS CONTROL LAYER PACKET DELAY DISTRIBUTION FOR IEEE 802.11

Hatm Alkadeki, Xingang Wang and Michael Odetayo

Department of Computing, Coventry University, Coventry City, UK

**ABSTRACT**

*The most important standard in wireless local area networks is IEEE 802.11. This is why much of the research work for the enhancement of wireless network is usually based on the behavior of IEEE 802.11 protocol. However, some of the ways in which IEEE 802.11 medium access control layer behaves is still unreliable to guarantee quality of service. For instance, medium access control layer packet delay, jitter and packet loss rate still remain a challenge. The main objective of this research is to propose an accurate estimation of the medium access control layer packet delay distribution for IEEE 802.11. This estimation considers the differences between busy probability and collision probability. These differences are employed to achieve a more accurate estimation. Finally, the proposed model and simulation are implemented and validated - using MATLAB program for the purpose of simulation, and Maple program to undertake the calculation of the equations.*

## 1. INTRODUCTION

The IEEE 802.11 WLANs standard often called Wireless Fidelity (Wi-Fi) networks provide flexible wireless access [1]. The standard is built on two specifications layers known as the Physical (PHY) layer and the Medium Access Control (MAC) layer. The MAC layer is a very important part for supporting quality of service (QoS), in particular to support the multimedia applications. Therefore, many researchers pay a great attention to the MAC layer so as to try and solve many of the problems which restrict QoS. These problems include: delay, packet loss rate, and jitter, which are still opening research questions. Due to the problems with the MAC layer the guaranteeing of QoS in IEEE 802.11 is still a very challenging task [2]. This is why QoS to IEEE 802.11 has become an active research area [2].

Estimating the MAC layer packet delay distribution will provide necessary information for QoS enhancement [3]. However, the primary mechanism for the MAC layer is based on Distributed Coordination Function (DCF) [4]. The DCF is based on Listen Before Talk (LBT) mechanism to detect whether a channel is idle or busy so as to avoid a collision. Moreover, the DCF uses four-way hand shaking (RTS/CTS-DATA/ACK). This mechanism helps the protocol to reduce number of collisions. These events will cause delay during a transmission process, which are considered in this paper. This paper aims to improve the work of [3] by considering the differences between the busy probability and the collision probability. Then, the accuracy of our new mathematical model is evaluated by carrying out simulation using the MATLAB program.

The rest of the paper is organized as follows. Section 2 summarizes and evaluates number of popular related work. Section 3 presents the MAC layer delay as a terminating renewal process. Section 4 presents the numerical calculations for our new mathematical model. Section 5 compares the results obtained from our model with the simulation results for wireless node behavior model based on the Bianchi's model. Moreover, the results will also be compared with the previous work. Section 6 concludes our study.

## 2. RELATED WORK

Most of the popular work for studying the behavior of a single node and performance for wireless network is based on the Markov chain probability model. The back-off mechanism for IEEE 802.11 DCF is represented by a bi-dimensional discrete time Markov chain [4]. Therefore, Bianchi in [4] proposed a good evaluation for IEEE 802.11 DCF performance but it has some limitations which need to be investigated. This is why a number of researchers are working on extending that model so as to improve the performance of wireless networks (WLANs). For instance, in [5], the authors improved [4] by using the same assumption to consider the effect of packet retransmission limit during their analysis but using a new scheme called DCF+ to improve the performance of TCP. This means that the authors worked on the MAC layer to improve the performance analysis and the transport layer to support the transmission of packets over WLANs.

On the other hand, many other researchers focused on the MAC layer delay rather than the transport layer. In [6], the authors developed the performance analysis model in [5] to calculate the packet delay. The authors proved that the result from their model is better than the result from the model without considering packet retry limits. However, the authors in [7], improved [4] from the bi-dimensional Markov chain to a single dimensional to compute throughput. On the other hand, they calculated the average packet delay by reducing the model in [5] from two to one dimension. However, one-dimensional Markov chain is a good idea for a simple calculation but it is not suitable for large network. Therefore, in [8], the authors proposed a new analytical model to calculate the average packet delay and improved the model in [6]. Furthermore, in [9], the authors evaluated that [7] is an inexact model for a small network which has less than 20 nodes. The authors proposed a new delay model by extending their previous work in [8]. Their study proposed most delay events as packet delay average, distribution, and jitter of IEEE 802.11 DCF. In [10], the authors estimated a delay, jitter, and throughput by using a multiplayer in a wireless network game model. On the other hand, in [11], the authors improve the work in [6] to study delay distribution of DCF. However, the study assumed the channel is idle, which means collision probability and busy probability are not included. Thus in [3], the authors proposed estimation for the MAC layer delay by using a mathematical model based on successful transmission. On the other hand, the model does not propose busy probability. Some of the researchers proposed something different to increase throughput and decrease mean packet delay for IEEE 802.11 DCF protocol. Therefore, in [12], the authors proposed a new back-off algorithm (EBA) to improve throughput by decreasing the delay packet.

As we have seen most existing models do not consider the differences between the busy probability and the collision probability. Therefore, in [13], the authors argued that the assumption that the busy probability and the collision probability are similar is not valid. Therefore, the delay occurred from frozen period in the busy state must be considered in mathematical models. Furthermore, most of all the above research work considered the estimations of the average MAC layer packet delay but the packet delay distribution is still unsolved [3].

# 3. ESTIMATION OF MAC LAYER PACKET DELAY DISTRIBUTION

In this paper, we consider all possible events during the back-off transmission mechanism for IEEE 802.11 DCF as shown in figure1. Therefore, the total time is regarded as a sequence of intervals of Empty ($D_{Emp}$), Successful ($D_{Suc}$), Busy ($D_{Bus}$) and Collision ($D_{Col}$) delay time.

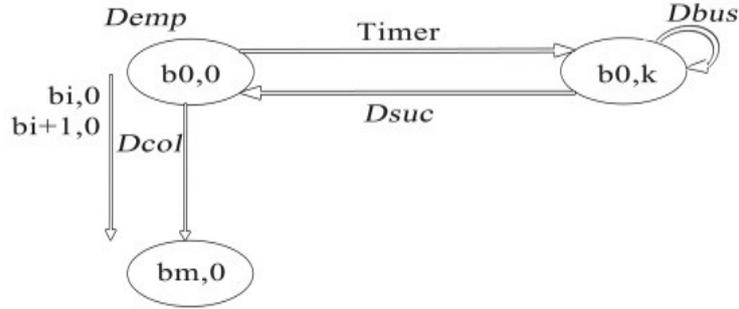

Figure1. Time Events during the back-off mechanism in IEEE 802.11 DCF represented by bi-dimensional Markov chain

Therefore, the time delays are calculated using the following equations:

$$D_{Emp} = 50 \ \mu s \qquad (1)$$

$$D_{Suc} = RTS + SIFS + CTS + SIFS + H + E[P] + SIFS + ACK + DIFS \qquad (2)$$

$$D_{Col} = RTS + DIFS \qquad (3)$$

$$D_{Bus} = DIFS + SIFS + ACK \qquad (4)$$

The network in this paper has 20-30 nodes; each node is equally likely to transmit. This uniform distribution considered because the time was very short. Therefore, this proposed model is based on two probabilities to represent the behavior model for each state. We supply $\tau_{tr}$ as the probability of the sender station attempting a transmission, $\tau_{nb}$ probability of one neighbor station attempting a transmission. In this case, we have the following possible different probability events, which we are able to calculate from (5), (6), (7), (8) and (9):

$P_{Emp}$ → No transmission attempts (idle)

$P_{Suc}$ → One of neighbor's attempting to transmit

$P_{Own}$ → The sender attempting to transmit

$P_{Col}$ → Transmission attempts at the same time

$P_{Bus}$ → Channel busy by transmission or collision

$$P_{Emp} = (1 - \tau_{tr}).(1 - \tau_{nb})^{n-1} \quad (5)$$

$$P_{Suc} = (n-1).\tau_{nb}.(1 - \tau_{tr}).(1 - \tau_{nb})^{n-2} \quad (6)$$

$$P_{Own} = \tau_{tr}.(1 - \tau_{nb})^{n-1} \quad (7)$$

$$P_{Col} = \tau_{tr}.(n-1).\tau_{nb}.(1 - \tau_{nb})^{n-2} \quad (8)$$

$$P_{Bus} = 1 - P_{Emp} - P_{Own} - P_{Suc} - P_{Col} \quad (9)$$

The time taken by a packet from its source to its destination is called delay [1], which means:

Total Delay= Delay on upper layer + Delay on MAC layer

In this study, we work only on the delay on the MAC layer. Therefore, we represent the delay on the MAC layer by the terminating the renewal process model to develop the work in [3]. In this case, the collision or busy will cause the time delay of Ti seconds. The delay is represented as sequence of discrete random variables and terminated by each successful transmission is as shown in (10).

$$S_n = T_1 + T_2 + \cdots + T_n + D_{Suc} \quad (10)$$

Ti: Represents discrete random variable for time delay in seconds when a station will face in case of a collision or frozen period. All Ti have the same improper probability distribution function (F) and probability density function (*f*).

In this paper the $D_{Emp}$, $D_{Suc}$, $D_{Col}$, $D_{Bus}$ are the random variables whose corresponding probability density functions are $P_{Emp}$, $P_{Suc}$, $P_{Col}$, $P_{Bus}$ which are obtained from (5), (6), (8) and (9). Therefore,

$$P_{Emp} = f(D_{Emp})$$
$$P_{Suc} = f(D_{Suc})$$
$$P_{Col} = f(D_{Col})$$
$$P_{Bus} = f(D_{Bus})$$

However, $P_{Emp}$, $P_{Suc}$, $P_{Col}$, $P_{Bus}$ present the probabilities of the slots or transmission attempts in which a station will not transmit. Therefore, probability distribution function (F) will be equal to 1 if $P_{Own}$ is added to it.

F defines in the paper as: $F(\infty) = 1 - P_{Own}$

From theory of probability and stochastic we know that the *f* can be obtained by taking the derivation of F. On the other hand, F can be obtained by integrating the *f*. Therefore, from the renewal process theory [14], if we compare the basic renewal equation with (11):

$$\int_0^\infty e^{x.y} . F(dy) = 1 \quad (11)$$

We realize x is the transform variable or the Laplace transform variable. Therefore, the (12) could be derived by obtaining the value of x. The process terminates after a time value t:

$$P(M > t) \approx \frac{1 - F(\infty)}{X.\mu} . e^{-x.t} \qquad (12)$$

Where

$$\mu = \int_0^\infty y. e^{x.y} . F(dy) \qquad (13)$$

We can consider (11) and (13) as the following:

$$P_{Emp}. e^{x.D_{Emp}} + P_{suc}. e^{x.D_{suc}} + P_{col}. e^{x.D_{col}} + P_{Bus}. e^{x.D_{Bus}} = 1 \qquad (14)$$

$$D_{Emp}. P_{Emp}. e^{x.D_{Emp}} + D_{Suc}. P_{suc}. e^{x.D_{suc}} + D_{Col}. P_{col}. e^{x.D_{col}} + D_{Bus}. P_{Bus}. e^{x.D_{Bus}} = \mu \qquad (15)$$

As a result of obtaining x value from (14) and μ value from (15), we can estimate (12). However, estimating P {M>t} allows us to estimate the MAC layer packet delay distribution for IEEE 802.11 as follows:

$$P\{d \in [a; b]\} = P\{M > a\} - P\{M > b\} \qquad (16)$$

$$P\{d \in [0; c]\} = 1 - P\{M > c\} \qquad (17)$$

Equations (16) and (17) represent the MAC layer delay distribution (d) as a histogram, which are based on estimating (12) after time value t.

## 4. NUMERICAL RESULT

The time delays were calculated with the help of (1), (2), (3), and (4) and the values given in Table1.

Table 1. System parameters.

| Parameter | Value |
|---|---|
| SIFS | 28μs |
| DIFS | 128μs |
| EIFS | 456μs |
| PHYSICAL SLOT | 50μs |
| RTS | 350μs |
| CTS | 350μs |
| ACK | 300μs |
| DATA PACKET | 8200μs |
| NETWORK NODES (n) | 20-30 nodes |
| Service Time (ms) | 0-200 ms |

Therefore, the time delay values are as follows:

$$D_{Emp} = 50 \mu s$$

$$D_{Suc} = 9412 \ \mu s$$

$$D_{Col} = 478 \ \mu s$$

$$D_{Bus} = 456 \ \mu s$$

The probabilities are calculated with the help of (6), (7), (8), and (9) where n=20. Therefore, the probabilities values are as follows:

$$P_{Emp} = 0.3585$$

$$P_{Suc} = 0.3585$$

$$P_{Col} = 0.0189$$

$$P_{Own} = 0.0189$$

$$P_{Bus} = 0.2453$$

Now, to obtain the real value of x we need to calculate the roots of (14). It will be easily done by converting it into polynomial equation and then we obtain a real root as shown in (18) and (19). Therefore, by substituting $e^{X.50.10^{-6}} = t$ and approximating (14) we obtain:

$$0.3585t + 0.2453t^9 + 0.0.0189t^{10} + 0.3585t^{188} = 1 \qquad (18)$$

$$0.3585t + 0.2453t^9 + 0.0.0189t^{10} + 0.3585t^{188} - 1 = 0 \qquad (19)$$

This equation has 188 possible solutions so we will use the Maple program for numerical calculation to find the roots of the above equation. The real root for t is 1.000261721.

However, $t = e^{X.50.10^{-6}}$ then,

$$1.000261721 = e^{X.50.10^{-6}}$$

$$\ln(1.000261721) = X.50.10^{-6}$$

$$X = 5.234$$

Therefore, by obtaining x value then we can obtain the value for μ from (15). Finally, we use the values for x and μ to solve (12) during the interval time between 0 and 200ms. The result produces a histogram between probability and time interval as shown in Figure2.

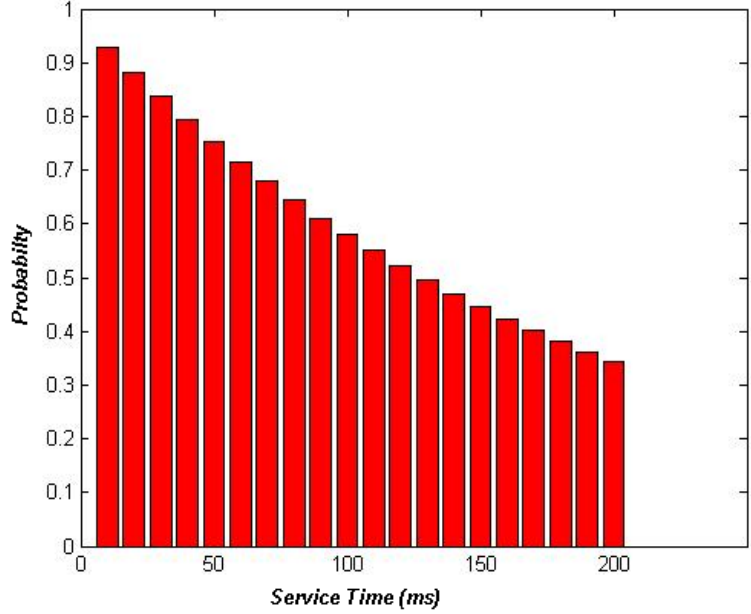

Figure2. Terminating renewal process model (n=20)

## 5. MODEL VALIDATION

In this section, we compare the result from our mathematical model with the wireless network behavior based on the Bianchi's model in [4]. Furthermore, the result will also be compared with the previous work in [3]. In this case, we use the same assumption as in the previous work for $\tau_{tr}$ and $\tau_{nb}$ both are considered to be the same as $\tau$ value in [4] as follows:

$$P = 1 - (1 - \tau)^{n-1} \qquad (20)$$

$$\tau = \frac{2}{1 + W + P.W.\sum_{i=0}^{m-1} 2xP^i} \qquad (21)$$

The system parameters used for both mathematical model and simulation are in Table1. In addition to the system parameters, $W$ denotes initial contention window size and m denotes maximum back-off stage. Figures3 and 4 provide us knowledge to possible estimation for packet delay distribution by using (16) and (17). As it can be seen from the figures, the results of the proposed model and the simulation are very close at each discrete time value. These results present for Packet Delay Right Tail Distribution Function (RTDF), where (RTDF(x) = P(X>x) for $x \in \mathcal{R}$ probability that packet delay exceeds x). In these experiments the errors do not exceed 0.0082 for networks of 20 nodes and 0.0025 for networks of 30 nodes. Moreover, these experiments show a good accuracy compared with previous work in [3], where the errors were 0.0332 for networks of 20 nodes and 0.0235 for networks of 30 nodes. On the other hand, we agree with the previous work in [3], where the effects for $D_{Emp}$, $D_{Col}$, $D_{Bus}$ are minor in comparison to that of $D_{Suc}$. Furthermore, our new mathematical model provides accurate way for estimating the MAC layer packet delay distribution.

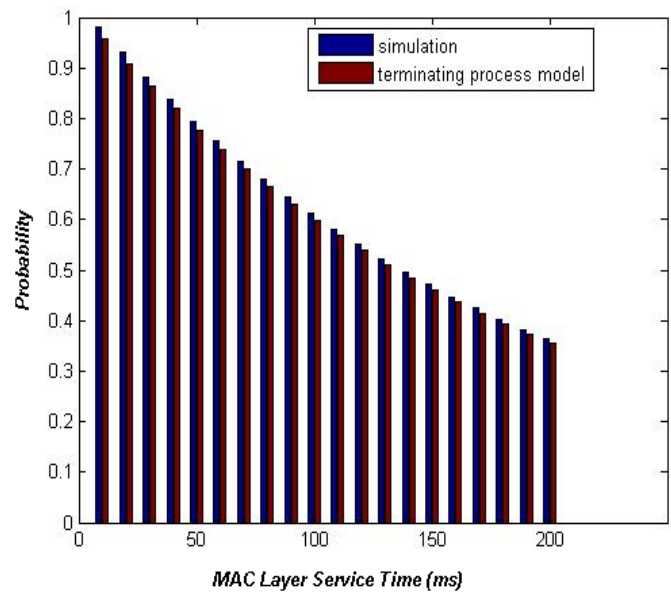

Figure3. Terminating renewal process model comparing with simulation (n=20).

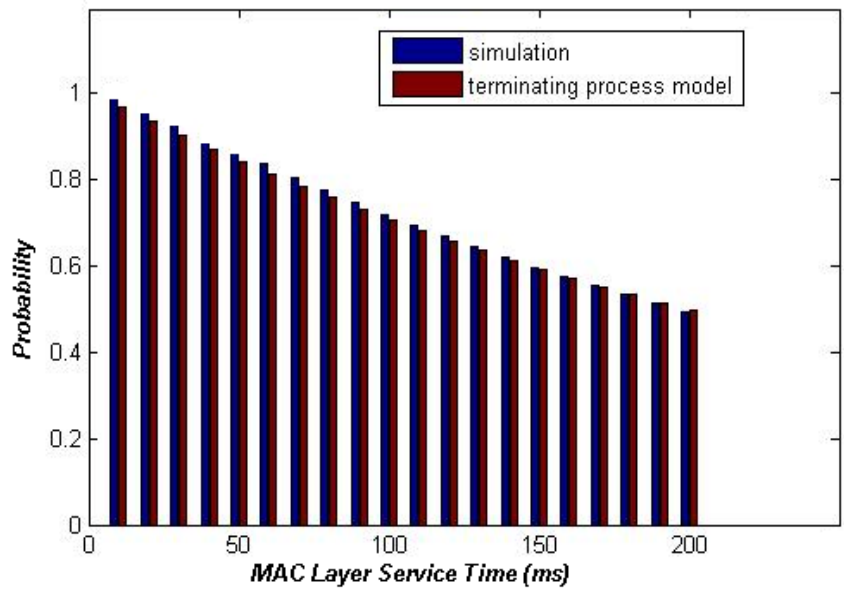

Figure4. Terminating renewal process model comparing with simulation (n=30).

## 6. CONCLUSIONS

Packet delay distribution depends on the MAC layer. The MAC layer provides a way for channel access. However, several events can happen during the channel access. Therefore, these events cause delay during transmission. The proposal uses the terminating renewal process theory for modeling MAC layer delay. The proposed solution considering the difference between the busy probability and the collision probability, which will lead to improve the accuracy for estimating the MAC layer packet delay distribution for single-hop wireless network.

This paper provides a good agreement between our mathematical model and wireless network behavior simulation. Therefore, the model provides prediction of high quality as expected.

In our future work, we will consider to use new system parameters where timeslots are five times shorter than Bianchi's model parameters. These will enable us to further develop our model with more realistic parameters.

## ACKNOWLEDGEMENTS

The authors wish to thank the anonymous reviewers for their useful comments that have significantly improved the quality of the presentation. This work has received support from the Ministry of Higher Education in Libya and Coventry University in UK.

**Authors**

**Mr. Hatm Alkadeki** is currently doing PhD research on Performance Enhancement for Wireless Network Protocol in Department of Computing and Digital Environment, Coventry University, UK. He graduated from Computer Engineering Department of Engineering Academy, Tajoura, Libya. He obtained MSc degree in Computer and Control Engineering from Budapest University of Technology and Economics, Budapest, Hungary. He has worked in Academy Tajoura as teaching assistant and lecturer. He also worked as lecturer at other universities in Tripoli. Hatm is also member in Engineering Association in Libya, Global Leaders Programme (GLP) and IEEE. 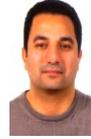

**Dr. Xingang Wang** is currently a senior lecturer in Department of Computing and Digital Environment, Coventry University, UK. He has worked in Centre for Security, Communications and Networks Research, University of Plymouth for 5 years as lecturer before joining the department. He received his PhD in performance modelling design and analysis of multiple access protocols and Quality of Service schemes for computer networks from University of Bradford, UK in 2005. He is current research interests include designing MAC protocols, integration of networks and enhancing performance of networks. He has held a number of grants and published over 50 publications in these areas. He is currently co-supervising a number of PhD students on various topics in these areas and is currently involved in one industrial collaboration project on Quality of Service management in packet switched networks. He has actively served the academic community and cofounded International Workshop on Performance Modeling and Evaluation in Computer and Telecommunication Networks. He has also acted as a co-chair for a number of international conferences and served on program committees for over 10 international conferences and served on program committee for over 10 international conference/workshops. He has edited a number of journal issues as a guest editor and been a referee for many international journals including IEEE Wireless Communications Magazine, Elsevier Journal on Computer Networks, Elsevier Journal on Wireless Communications and ACM/Kluwer Mobile Networks and Applications Journal. Dr Wang also holds the Microsoft Certified System Engineer (MCSE) professional qualification and is a member of IEEE and IET. 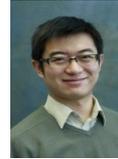

**Dr. Michael Odetayo** is a principal lecturer in the Department of Computing at Coventry University, UK. He obtained his MSc degree from the Department of Computer Science, Imperial College, London, UK and his PhD Degree in ComputerScience at the University of Strathclyde, Glasgow, UK. Before completing his PhD, he worked at the Computer Centre of Ahmadu Bello University, Zaria, Nigeria, where he led many project teams that designed and developed computer basedsystems for the University and other organisations outside the University. Hejoined De Montfort University, Leicester, UK, after completing his PhD where he was a senior lecturer in the Department of Computer Science for many years before moving to Coventry University. His research areas include Genetic Algorithms, Hybrid Systems, Biomedical Intelligent Systems, Intelligent Knowledge-Based Systems, e-Commerce, Neural Networks, Image enhancement for Biometric security systems, Data Mining and Machine Learning. 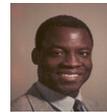